\documentstyle[preprint,aps]{revtex}

\begin{document}
\draft
%
%
\title
{Nonlinear magneto-optical Kerr spectra
of thin ferromagnetic iron films calculated with {\em ab initio} theory}
\author{U. Pustogowa, W. H\"ubner, and K. H. Bennemann}
\address{Institute for Theoretical Physics,
Freie Universit\"at Berlin, Arnimallee 14, D - 14195 Berlin, Germany}
\author{T. Kraft}
\address{Fritz-Haber-Institut der Max-Planck-Gesellschaft, Faradayweg 4-6,
D - 14195 Berlin, Germany}
\maketitle
%
%
\begin{abstract}
Using a
spin-polarized full-potential linear muffin-tin orbital method
we present calculations of the nonlinear magneto-optical Kerr effect
for thin {\em bcc} Fe films within a slab geometry.
Film layer dependent contributions to the Kerr spectrum are determined.
Thus, we calculate the magneto-optical Kerr spectra for the linear and
nonlinear case.
Our results show clearly that the Kerr spectra of thin films are
characteristicly different from those at surfaces of bulk materials.
In the case of linear Kerr spectra of Au/Fe({\em bcc})/Au(001) films
our theoretical results are in good agreement with observed
frequency- and thickness-dependent spectra.
\end{abstract}

\newpage
%
\section{INTRODUCTION}
The nonlinear magneto-optical Kerr effect (NOLIMOKE) has been proven to be a
sensitive probe for the electronic, magnetic, and symmetry properties
of ferromagnetic surfaces and to yield new results.
For example, it was shown theoretically~\cite{raps} and
experimentally~\cite{Reif,Reifrau,RasKo,Boeh},
that the nonlinear Kerr rotation angle
is significantly enhanced compared to the linear Kerr angle
and furthermore that the Kerr spectra exhibit characteristic features of the
bandstructure~\cite{Pusto1}.
Thus, one expects that NOLIMOKE will be also a successful tool to study
magnetic multilayers with contributions of every interface
within the light penetration depth.
Actually, experiments at a Co/Au multilayer system~\cite{Rasprb} show the
dependence of the reflected second-harmonic intensity on the number of
interfaces.
In view of the surface sensitivity of NOLIMOKE it is necessary for
comparison of theory with experiments to perform electronic calculations
for NOLIMOKE in thin films.
We show that it will not be sufficient to simulate films by simply scaling
calculations of bulk surfaces.
Actually, linear surface MOKE experiments on Au/Fe({\em bcc})/Au(001)
observe new structures in the frequency-dependent Kerr spectrum for thin
Fe films~\cite{Katayama1}.
Possibly, even quantum-well states are reflected in recent experiments
on this system~\cite{Katayama2}.

In view of this we performed the first {\em ab initio} calculations of
magneto-optical Kerr spetra for {\em bcc} Fe films with varying thickness.
The film bandstructure is calculated using the
spin-polarized full-potential linear muffin-tin orbital method (FP-LMTO).
For comparison with the experiment we performed all calculations for
nonlinear and linear Kerr spectra.
Note, in our calculations we use ${\bf k}$-independent transition matrix
elements and thus we have to implement the surface sensitivity.
For this we use two models:
(i) We take the first layer as an effective response volume.
(ii) The surface response is calculated directly by
projecting the wave functions to the top film layer.
Hence we assume that the nonlinear response results only from the
surface layer.
The nonlinear response senses only the electronic structure of the first
two top film layers, whereas in the case of linear response
the electronic structure of the whole film is involved.
This is found to be in agreement with
experiments~\cite{Wuttig,Bader,RasKir,Rasprl}.

To simulate structural film effects we perform calculations for different
film lattice constants.

In order to check the validity of using and scaling bulk parameters for film
calculations, we compare the monolayer and the bulk {\em ab initio}
calculated spectra with results of semi-empirical tight-binding calculations
using scaled parameters to simulate the reduced dimension.
The semi-empirical monolayer spectrum is calculated firstly on the basis of
a three-dimensional bandstructure with bulk parameters scaled with the
square root of the coordination number
and secondly using a two-dimensional bandstructure.
We find that scaled bulk parameters are not sufficient to characterize
film spectra properly.

For the two-dimensional semi-empirical calculations, spin-orbit coupling
is treated using two different approximations:
First, we calculate spin-orbit coupling non-perturbatively in the
Hamiltonian.
Secondly, we employ first-order perturbation theory.
Thus, we are able to prove the linear dependence of the nonlinear
Kerr effect on spin-orbit coupling.

In Section II we outline our electronic theory for the film calculations,
in Section III we present results and these are discussed in Section IV.
Summarizing, our results show that:
(i) Nonlinear Kerr spectra reflect the film structure
more sensitively than linear ones.
(ii) Nonlinear Kerr spectra reveal special features of films and monolayers
which are different from those of bulk surfaces.
Our results indicate a magnetic moment enhancement at
surfaces and interfaces and in thin films.
%

%
\section{THEORY}
The nonlinear magneto-optical surface susceptibility
$\chi^{(2)}(2q_{\parallel },2\omega ,{\bf M})$
was derived within an microscopic and electronic theory by H\"ubner
{\em et al.}~\cite{Hue89,Hue90}.
The calculations of the nonlinear magneto-optical Kerr spectra of
Ni and Fe surfaces of bulk crystals~\cite{Pusto1} use the
following expression
\begin{eqnarray}
    &&\chi^{(2)}_{xzz}(2q_{\parallel },2\omega ,{\bf M})=
    \frac{e^{3}C}{\Omega} \frac{\lambda_{s.o.}}{\hbar \omega } \nonumber\\
    &&\times \sum_{\sigma }\sum_{{\bf k},l,l^{\prime },l^{\prime \prime }}
	\Bigg\{
    \langle {\bf k}+2{\bf q}_{\parallel },l^{\prime \prime }\sigma |z|
        {\bf k}l\sigma \rangle
      \langle {\bf k}l\sigma |z|{\bf k}+{\bf q}_{\parallel},l^{\prime }\sigma
        \rangle
      \langle {\bf k}+{\bf q}_{\parallel},l^{\prime }\sigma |z|
        {\bf k}+2{\bf q}_{\parallel },l^{\prime \prime }\sigma \rangle
     \nonumber\\
    &&\times  \frac
    {\frac{f(E_{{\bf k}+2{\bf q}_{\parallel },l^{\prime \prime }\sigma})-
           f(E_{{\bf k}+{\bf q}_{\parallel },l^{\prime }\sigma })}
         {E_{{\bf k}+2{\bf q}_{\parallel },l^{\prime \prime }\sigma }-
          E_{{\bf k}+{\bf q}_{\parallel },l^{\prime }\sigma }
          -\hbar \omega + i\hbar \alpha_{1}}
    -\frac{f(E_{{\bf k}+{\bf q}_{\parallel },l^{\prime }\sigma })-
           f(E_{{\bf k}l\sigma })}
         {E_{{\bf k}+{\bf q}_{\parallel },l^{\prime }\sigma }-
          E_{{\bf k}l\sigma }-\hbar \omega +i\hbar \alpha_{1}} }
    {E_{{\bf k}+2{\bf q}_{\parallel },l^{\prime \prime }\sigma }-
     E_{{\bf k}l\sigma }-2\hbar \omega +i2\hbar \alpha_{1}}
	\Bigg\} \; ,
\label{gl1}
\end{eqnarray}
where $\lambda_{s.o.}$ is the spin-orbit coupling constant,
$E_{{\bf k}l\sigma }$ are the electronic energy levels resulting from the
bandstructure calculations, and the factor $C$ determines the surface
treatment as discussed below.
The other symbols are the same as in previous studies (e.g.~\cite{Pusto1}).
Here, we present the susceptibility tensor element $\chi_{xzz} $ as a
typical odd element in the longitudinal Kerr geometry ($M$ parallel
to the $y$ axis, $p$-polarized incident electric field).
Other magnetic (odd) and nonmagnetic (even) tensor elements can be written
similarly.
This formula applies also to films.
Thus, the bandstructure $E_{{\bf k}l\sigma }$ and the transition matrix
elements are calculated for films.
For the film calculation we use wave functions in the transition matrix
elements which depend on the position of the appropriate atom in the film.
This is necessary for determining layer-dependent contributions.

The surface sensitivity of the nonlinear Kerr effect and thus $\chi^{(2)}$
is controlled by the breakdown of inversion symmetry and is essentially
determined by the three transition matrix elements.
These sense the necessary breakdown of inversion symmetry.
Previous calculations of nonlinear and linear Kerr spectra of Ni have
shown that the position of the main structures in $\chi^{(2)}$
and the peak height ratio of $\chi^{(2)}$ do not depend on the $k$
dependence of the transition matrix elements~\cite{Hue90}.
Mainly, the energy eigenvalues define the spectral structure.
This was checked by comparison of MOKE calculations and experimental
obtained spectra.
Thus, we use in our calculations constant transition matrix elements
and simulate the surface sensitivity by an additional factor $C$:
(i) $C=q_{\parallel }a$ in the case of the surface of bulk material,
where $q_{\parallel}$ is the component of the photon wave vector parallel
to the surface and $a$ is the lattice constant.
This factor gives the ratio of the response depth $a$ to the
excitation depth in the crystal ($\approx 1/q_{\parallel}$).
(ii) In the case of films we use
(a) $C=1/n^{3}$, with $n$ atomic layers in the film and alternatively
(b) $C=W_{{\bf k}+2{\bf q}_{\parallel},l^{\prime \prime} \sigma }
W_{{\bf k}+{\bf q}_{\parallel },l^{\prime }\sigma } W_{{\bf k}l\sigma }$,
where $W_{\alpha }$ denotes the weight of the density of state
$|{\bf k}l\sigma \rangle $ in the Wigner-Seitz cell of the first
monolayer.
The expression (a) follows from taking into account only the surface
monolayer response whereby a film averaged electronic structure is used.
The factor $1/n^{3}$ results from the fact that all three states
involved in the optical transition must be localized in the first layer.
(b) The factor  $C=W_{{\bf k}+2{\bf q}_{\parallel},l^{\prime \prime} \sigma }
W_{{\bf k}+{\bf q}_{\parallel },l^{\prime }\sigma } W_{{\bf k}l\sigma }$
results from the projection of the six wavefunctions in the three
matrix elements to atoms in the first layer.
Obviously this will be the better approximation.

The most important input for the calculation of the nonlinear susceptibility
is the electronic bandstructure.
For thin Fe films we calculate the bandstructure within the
full-potential linear muffin-tin orbital method (FP-LMTO)
developed by M. Methfessel {\em et al.}~\cite{Meth}
in the spin-polarized version by M. van Schilfgaarde, adapted by one of the
present authors (T. K.) for {\em fcc} and {\em bcc} Fe films
on several substrates~\cite{Kraft}.
Thus, we combine the advantages of a parameter-free density-functional
method with the possibility of consideration of many $k$ points in the
Brillouin zone which are required for describing optical processes
(transitions).
Besides, the full-potential version describes more efficiently the
interstitial-induced properties in low dimension compared to the atomic
sphere approximation successfully used in bulk systems.
The local spin-density approximation (LSDA) is used with the
exchange-correlation functional in the Wosko-Wilk-Nusair
parametrization~\cite{Vosko} of the Ceperley-Alder results~\cite{Alder}
within the spin-density-functional formalism.
The basis set consists of Hankel functions for the $s$, $p$, and $d$
electrons augmented to numerical solutions of the radial Schr\"odinger
equation at three different energies.
In the interstitial region the potential and the charge density
are represented by linear combinations of Hankel functions
fitted to the values and slopes on the atomic spheres.
For the calculation of the nonlinear Kerr spectrum we extract after the
self-consistency cycle the eigenvalues and the
eigenfunctions at all $k$-points.
We performed calculations of free-standing {\em bcc} iron films
within a slab geometry~\cite{Duedol}.

This type of {\em ab initio} calculation permits now to determine the
thickness dependence of the nonlinear Kerr spectrum including characteristic
features of the electronic and geometric structure of films.
It is of interest to compare the {\em ab initio} calculated results with
those obtained from tight-binding calculations
to shed more light on approximations.
Thus, we present results for the surface of bulk material using the
semi-empirical interpolation scheme as in previous
studies~\cite{Pusto1,Hue90}.
In addition, we calculate the Kerr spectra of films by simply scaling
the $d$-electron bandwidth of the bulk according to
$W \sim \sqrt{Z} A$, where $Z$ refers to an effective coordination
number and $A$ is the hopping integral.
This simulates the change in the coordination number due to the film
dimension.
The $s$-band width is kept unchanged, since mainly $d$ electrons contribute
to $\chi^{(2)}$ in the low frequency range.
In particular, this is a very rough approximation for MOKE.

Finally, we have performed a direct (unscaled) two-dimensional
tight-binding calculation of a monolayer for $d$ electrons only
using a quadratic lattice to which {\em fcc} and {\em bcc} structures
simplify in two dimensions.
{}From these various calculations we will learn about the necessity to
perform {\em ab initio} calculations rather than tight-binding calculations
to determine the characteristic features of the thin-film Kerr spectra.

To shed light on structural film effects, we have performed {\em ab initio}
calculations of nonlinear Kerr spectra for Fe monolayers with different
lattice constants.

To check the perturbative treatment of spin-orbit coupling resulting from
the wave functions [see  Eq. (~\ref{gl1})] we have also calculated
$\chi^{(2)} $ by including exactly the spin-orbit coupling
in the energy eigenvalues.
The spin-orbit dependent Hamiltonian is used in the form of Bennett {\em et
al.}~\cite{Bennett}.
Since in iron the majority and minority pure $d$ bands are separated by a
large exchange splitting of $J_{0}$=1.78 eV, the spin-orbit coupling
cannot mix the spin up and spin down bands and the only influence
of spin-orbit interaction on the energy bands in the Fe $d$ bands
appears at points of band crossing where the degeneracy of the bands is
lifted and the crossing points transform into extrema~\cite{Moos}.
%

%
%
\section{RESULTS}
In Figs. 1 and 2 results are shown for the thickness dependence of
the magneto-optical Kerr spectra of {\em bcc} Fe films obtained
from {\em ab initio} calculations.
The film specific features are recognized by comparing the results for
the surface of corresponding bulk materials.
In Fig. 1(a) the surface sensitivity was simulated by using the factor
$1/n^{3}$.
Note, the differences between the nonlinear and linear spectra.
In the nonlinear monolayer spectrum the first minimum is shifted compared
to the surface of bulk spectrum by about 2 eV to lower energies.
This results from the reduced $d$-band width of a film.
Furthermore, the tiny maximum at 5.6 eV in the surface of bulk spectrum
gets enhanced and shifted to 4.5 eV, having then a width of nearly 3 eV
in the case of a monolayer spectrum.
In the nonlinear case the position of the minimum between 2 and 4 eV
and the slope up to 5.5 eV in the 7 layer spectrum lie close to the
surface of bulk spectrum.
Note, consistent with our approximation regarding the surface sensitivity,
both film results and the surface of bulk spectrum converge.
The first minimum around 1 to 2 eV is deepest for a monolayer.

In Fig. 2 the nonlinear spectra where obtained by simulating the
surface sensitivity using the approximation (b)
$C=W_{{\bf k}+2{\bf q}_{\parallel},l^{\prime \prime} \sigma }
W_{{\bf k}+{\bf q}_{\parallel },l^{\prime }\sigma } W_{{\bf k}l\sigma }$,
which is the better approximation for the nonlinear case.
Consistent with this approximation the structure changes not too much
going from three to five to seven monolayers.
Thus, the essential contribution to NOLIMOKE results from the surface layer,
however, the electronic structure of this layer is affected
by the other layers due to hybridization and next-nearest neighbor
interaction.
This point is corroborated by the results for the monolayer.
The results are different from those for the surface of bulk spectrum,
where we have used the truncated bulk approximation.
One should also note the layer dependent change of the depth of the first
minimum which is related to the value of the magnetization in the top layer.

In Fig. 1(b) we also present results for the linear Kerr effect
in order to compare with the nonlinear case and with recent experiments.
The amplitudes in the linear spectrum are scaled with
the film thickness, e.g. by $1/n^{2}$.
Without such scaling the spectra will grow with increasing film thickness.
Note, the film and surface of bulk spectra converge as it should be.
The linear spectra of the 7 layer film and of the truncated bulk do not
differ up to 5 eV.
In the linear monolayer spectrum the transitions between majority spin
electron bands begin to dominate at energies higher 3.3 eV (sign change).
In the optical range we find a new structure at 5.2 eV.

In agreement with experiment~\cite{Katayama1,Katayama2} we observe the shift
of the first peak with respect to the one and three monolayer spectra.
It is interesting that we do not obtain the extra peak in the linear
spectrum of thin films.
This supports the interpretation of this peak as resulting of
quantum-well-state transitions.

In Fig. 3 we show results for the nonlinear magneto-optical Kerr spectra for
a Fe monolayer assuming different lattice constants to simulate
film-structure effects.
We compare spectra obtained using the Fe bulk lattice constant $a$=2.76 \AA
$\, $ (determined by total energy minimization),
using $a$=2.776 \AA $\, $ corresponding to a Fe monolayer on Au, and
$a$=2.783 \AA $\, $ corresponding to Fe on Ag,
and using the experimental Fe lattice constant $a$=2.88 \AA .
The values for Au and Ag follow from the theoretical lattice constant
applying the experimental lattice mismatch.
Whereas the amplitude of the general minimum at about 1.5 eV shows no
clear dependence on the lattice constant, the energy for which
$Im \chi^{(2)}$ changes sign corresponds quantitatively to the ratios of
the lattice constants.
Thus, we demonstrate that NOLIMOKE reflects sensitively structural changes.

To gain more information about the possibility of using
tight-binding methods for treating thin films we compare with
{\em ab initio} results in Fig. 4.
First, note that the {\em ab initio} and the semi-empirical tight-binding
spectra of the bulk surface agree rather well in the optical range.
However, the calculations for the monolayer using {\em ab initio} or
different tight-binding methods differ drastically.
Curve 5 is obtained by including in the hopping parameters
correlations effects and the reduced number of nearest neighbors of a
monolayer.
The situation is similar in the linear case, see Fig. 4(b).

This comparison illustrates that bulk tight-binding parameters are
acceptable for the Kerr spectra at the bulk surface but not at films.
Here, a tight-binding method may be successful using specific parameters
for every film layer.
If appropriate film parameters are unavailable {\em ab initio} calculations
are mandatory to obtain the specific film features.

In Fig. 5 we demonstrate the effects resulting from including
non-perturbatively the spin-orbit coupling also in the electronic
energies $E_{{\bf k}l\sigma }$.
Since in the case of Fe spin-orbit coupling does not mix the minority and
majority spin bands we find a negligible effect.
%

%
\section{DISCUSSION}
We learn from our results obtained by an electronic theory
that NOLIMOKE reflects sensitively the electronic structure of thin films.
In contrast to the MOKE signal the NOLIMOKE signal originates for flat
surfaces essentially from the surface layer~\cite{fuss}.
However, the electronic structure of this layer depends on the film
thickness.
This can be clearly seen in our results.
Very interesting are also the results for Fe on a Ag and Au substrate,
simulated by the lattice constant of the Fe monolayer,
since they demonstrate that even structural effects are reflected in
the NOLIMOKE spectra.
This is of general interest regarding film growth and structural changes
occurring as a function of film thickness.

We emphasize that characteristic features of magnetism in thin films
like changes of the magnitude of the magnetic moments and of the
magnetization can be seen in the NOLIMOKE spectra.
Particularly, this can be seen from the depth of the minimum in $\chi^{(2)}$
around 3 eV.
The enhanced minimum results from an increase of the magnetic moments.
Actually, we obtain an enhanced moment of 2.8 $\mu_{B}$ in the first
film layer.
Since the magnitude of the magnetic moment is a fingerprint of the geometric
structure this can be used to determine the film geometry.
While MOKE already reflects characteristic film-averaged features,
NOLIMOKE clearly exhibits further interesting details.
Both MOKE and NOLIMOKE are suitable to study thin films in a
material-specific way, however only NOLIMOKE can be used to study
interface effects for which we expect similar features as for the surface.
Our results (see Figs. 5a and 5b) support once again that it is sufficient
for transition metals to treat spin-orbit coupling by non-degenerate linear
perturbation theory applied to the wave functions.
This is physically expected~\cite{Hue90}.

The comparison of the {\em ab initio} results with
semi-empirical tight-binding calculations shows that it is necessary to use
{\em ab initio} calculations for the films if only bulk tight-binding
parameters are available.
However, tight-binding calculations may be adequate if one uses as input
parameters those determined from {\em ab initio} calculations for thin films.

Concerning comparison with experiment it has been observed in agreement
with our theory that the MOKE signal increases linear with the
film thickness up to 20 monolayers whereas the NOLIMOKE signal remains
nearly constant.
Our theory suggests that the experimental enhancement of the NOLIMOKE
signal for 3 and 4 layer films at a frequency of 1.55 eV (corresponding to a
wavelength of 800 nm, see Fig. 5 in ~\cite{RasKir}) results from the
electronic structure and in particular from the enhancement of the
surface magnetization.
Furthermore, the experiments describing the thickness dependence of the
MOKE or NOLIMOKE signals are usually performed at one fixed frequency.
Our calculations have shown that for very thin films structures of the Kerr
spectra are shifted due to changes in the electronic structure of the films.
Thus, the Kerr susceptibility at a fixed frequency does not show a
monotonous increase with increasing film thickness depending on the choise
of the frequency respective to the spectrum structure.

The onset of the NOLIMOKE signal due to the appearance of magnetism
will reflect the structure of the surface.
For flat surfaces, the slope of the signal should be much larger
than for corrugated surfaces.
For film thickness larger than a few atomic layers the NOLIMOKE signal
will remain nearly constant, while the MOKE signal increases with film
thickness.
An enhanced magnetization in thin films is reflected by an enhanced
NOLIMOKE signal.

Regarding the enhancement of magnetic moments in thin films we deduce from
our NOLIMOKE spectra for a three layer thick film $\mu_{surface}=$ 2.64
$\mu_{B}$.
{}From our electronic bandstructure we obtain directly a film averaged
magnetic moment of 2.62 $\mu_{B}$.
{}From polarized neutron reflection for a 5.5 monolayer Fe film on Ag(001)
covered by Ag Bland {\em et al.}~\cite{Heinrich} deduced an averaged
magnetic moment of 2.58 $\mu_{B}$.
Here again the comparison of MOKE and NOLIMOKE spectra can clear up
whether the magnetic moments are enhanced (at the film-substrate interface
or at the film-vacuum interface, or at both).

Since we present also the first theoretical results of the thickness
dependence of the linear magneto-optical Kerr spectra we compare these with
recent experiments by Suzuki {\em et al.}~\cite{Katayama1} and Geerts
{\em et al.}~\cite{Katayama2}.
In agreement with these experimental results we obtain the shift of the
first minimum in $\chi^{(1)}$ dependent on the film thickness
and a quick convergence towards bulk results for more than four layers.
Note, we have not included the confinement of the substrate electrons
and thus quantum-well-states effects are not included in the calculation.

For future studies on NOLIMOKE it will be interesting to extend calculations
to multilayer systems and to analyze in more detail interface contributions
and lateral resolution.
A first principle evaluation of SHG should show that essentially only
surface and interface layers contribute and that second layer contributions
are of less importance, since the breakdown of inversion symmetry is not
as strongly felt as in the interface layer.
Furthermore, it is possible to extract the magnetic easy axis at
interfaces from NOLIMOKE~\cite{Hueneu}.
In particular, we will extend our NOLIMOKE calculations to analyze
{\em bcc} vs. {\em fcc} Fe structure on a substrate during film growth.
%

%
\newpage
\newpage
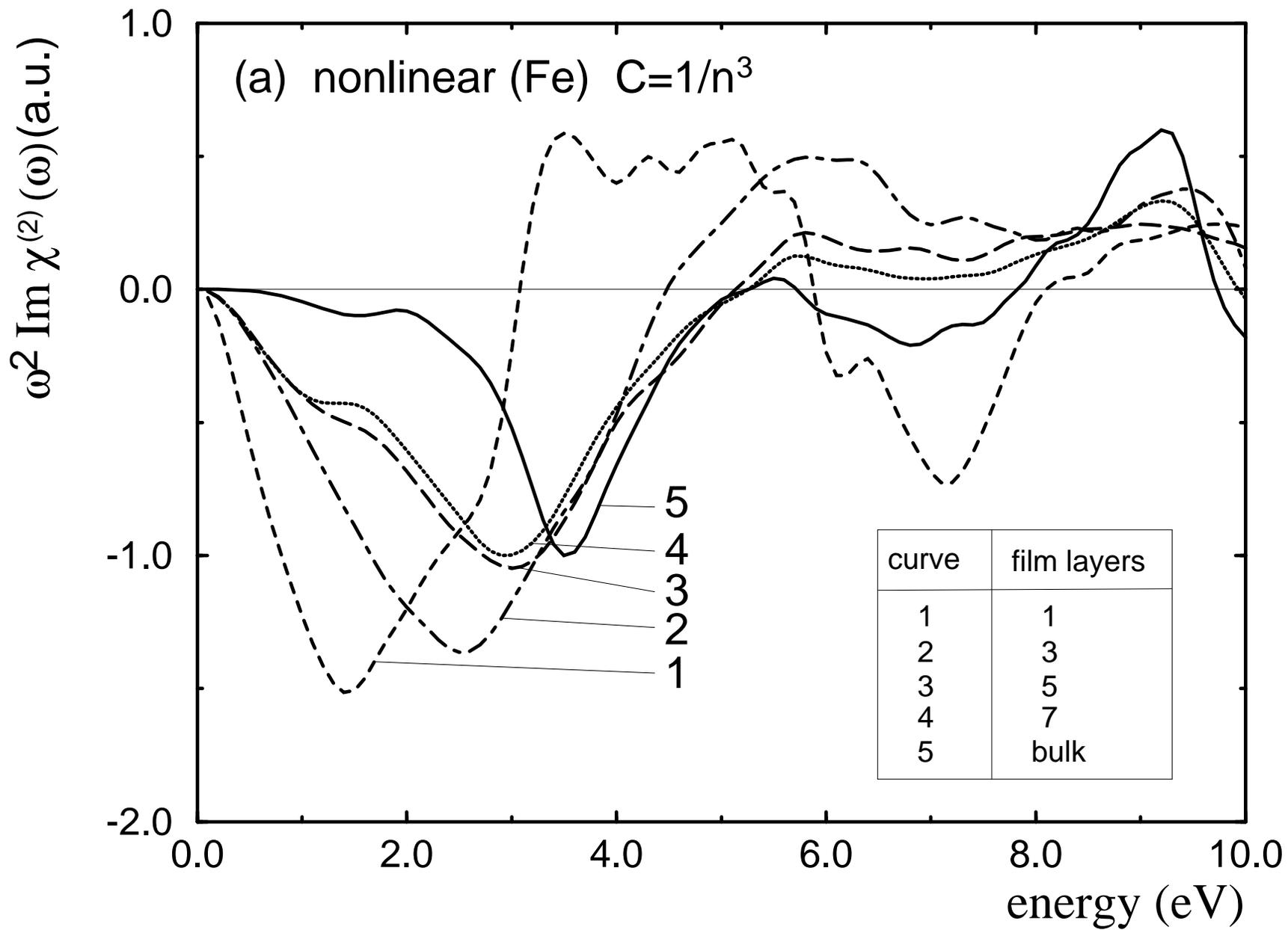
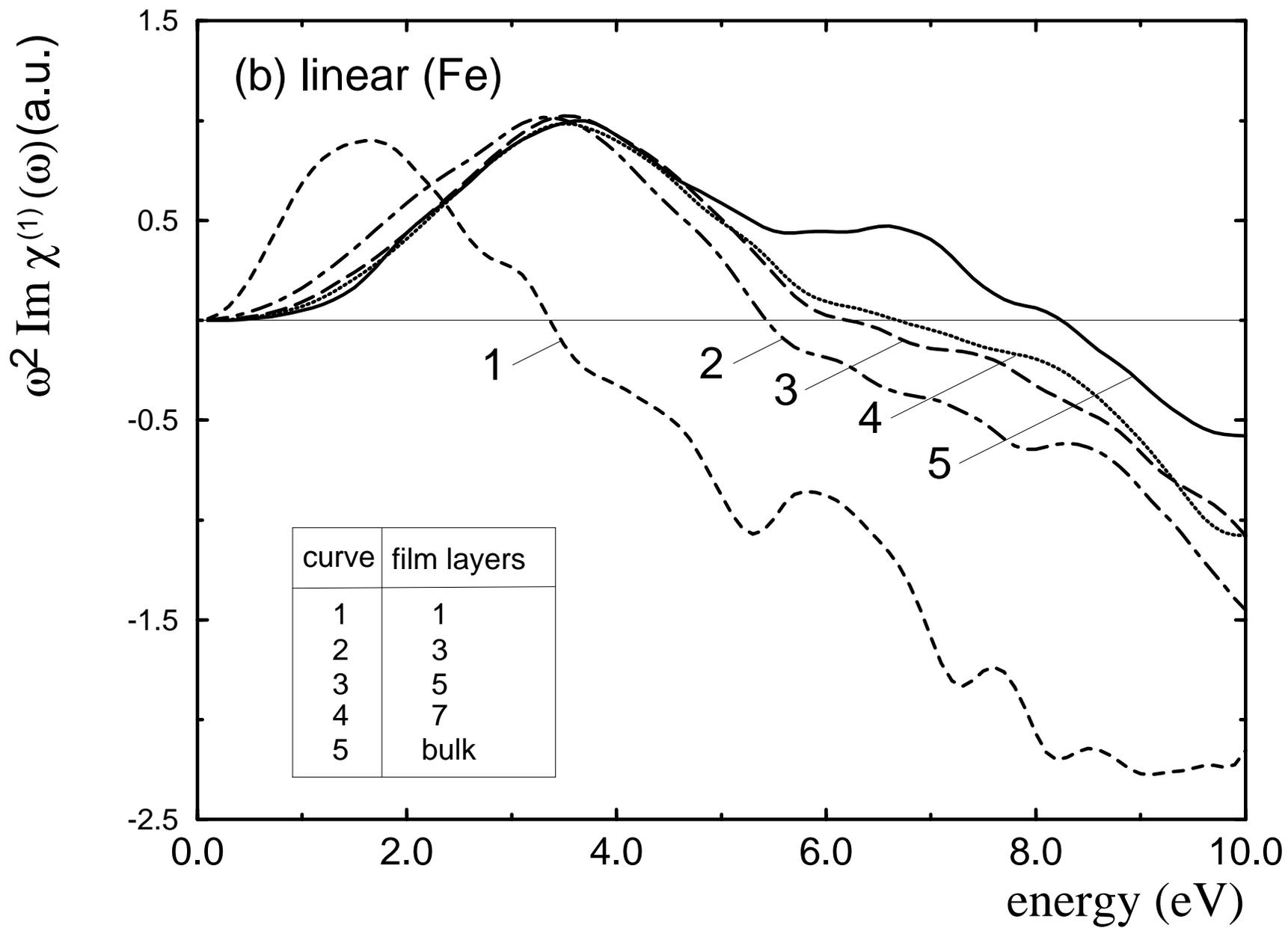
\begin{figure}
\caption[]{{\em Ab initio} calculated (a) nonlinear and (b) linear
magneto-optical Kerr spectra of Fe for a truncated bulk surface (solid
line), a monolayer (dashed curve), films having 3 layers (dashed-dotted),
5 layers (long-dashed), and 7 atomic layers (dotted curve).
The second harmonic response of the surface layer results by averaging
the electronic input structure over the whole film.}
\end{figure}
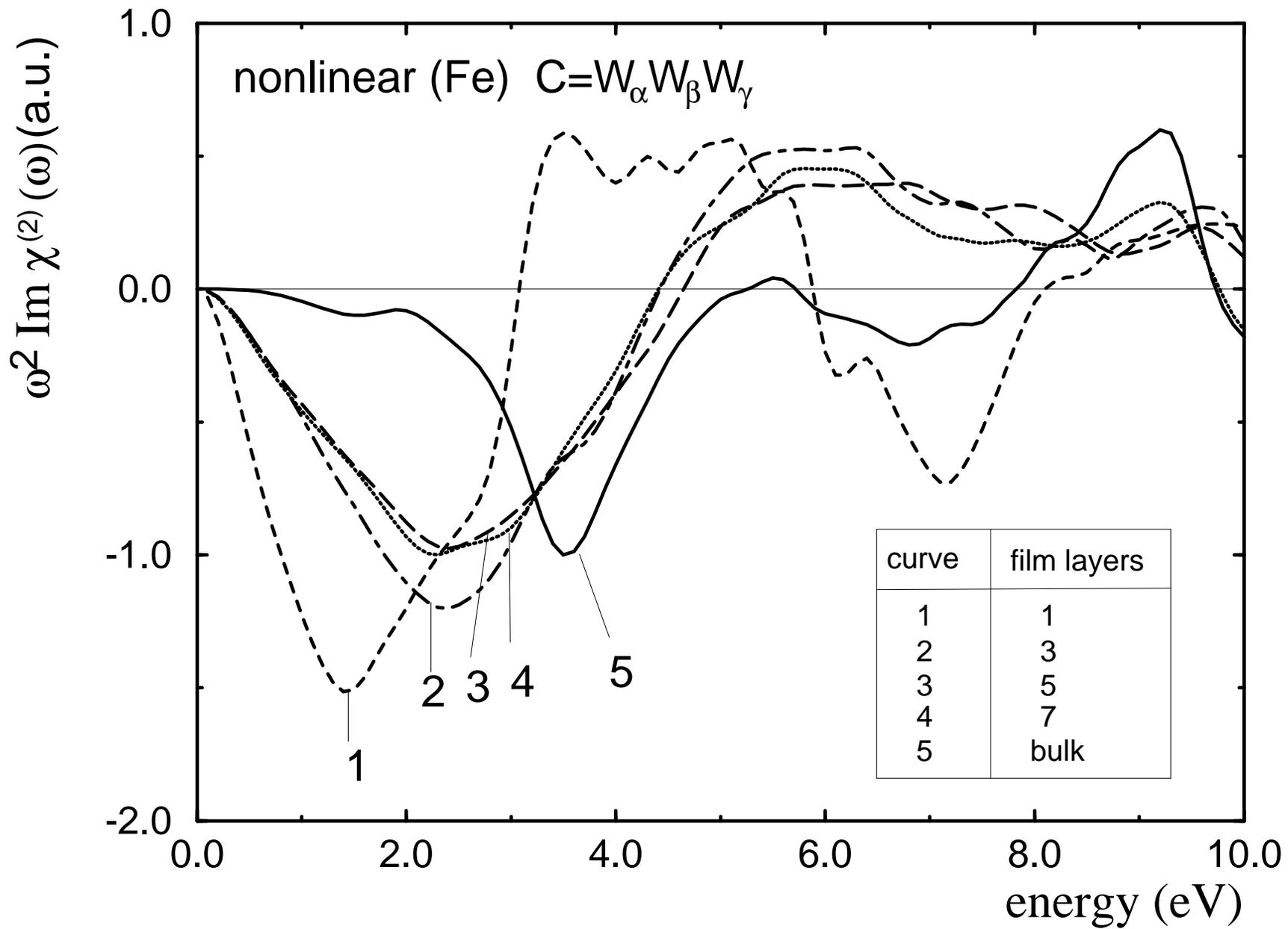
\begin{figure}
\caption[]{{\em Ab initio} calculated nonlinear magneto-optical Kerr spectra
of Fe for a truncated bulk surface (solid line), a monolayer (dashed curve),
3 layers film (dashed-dotted), 5 layers (long-dashed), and 7 atomic layers
(dotted curve).
The second harmonic response results from the first atomic layer.
This SH response is obtained by projecting the wave functions
to the first atomic layer yielding the factor $C$.}
\end{figure}
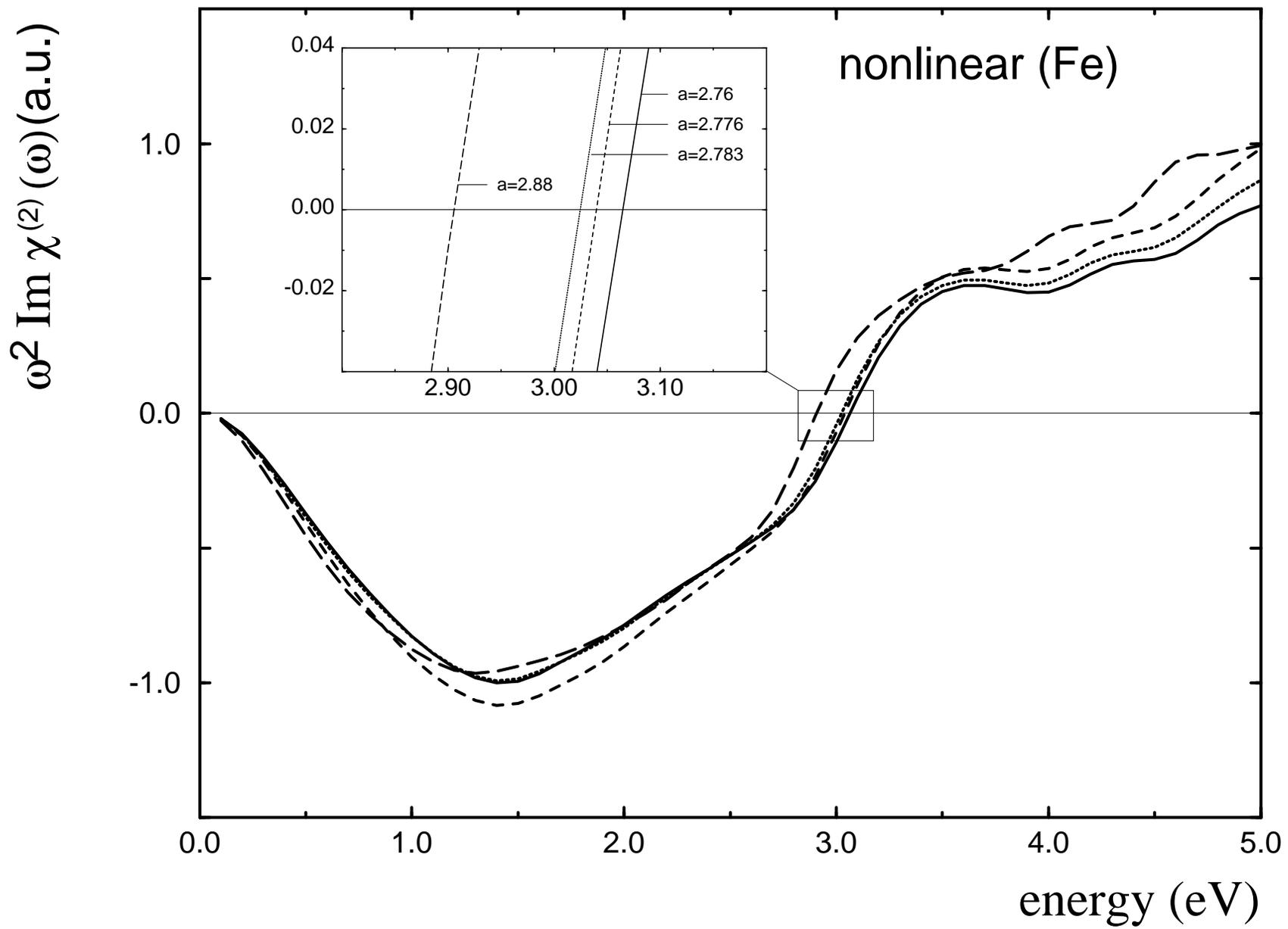
\begin{figure}
\caption[]{Film-lattice-constant dependence of {\em ab initio} calculated
nonlinear Kerr spectra of a Fe monolayer.
The solid curve refers to the bulk {\em bcc} Fe lattice constant, the dashed
curve to $a$=2.776 \AA (bulk Au), and the dotted curve to
$a$=2.783 \AA (bulk Ag).
The long-dashed curve refers to the experimental $a$ for Fe.
The inset shows at an enhanced scale the effects of different lattice
constants for the zero of Im $\chi^{(2)} $ at $\hbar \omega \approx $ 3 eV.}
\end{figure}
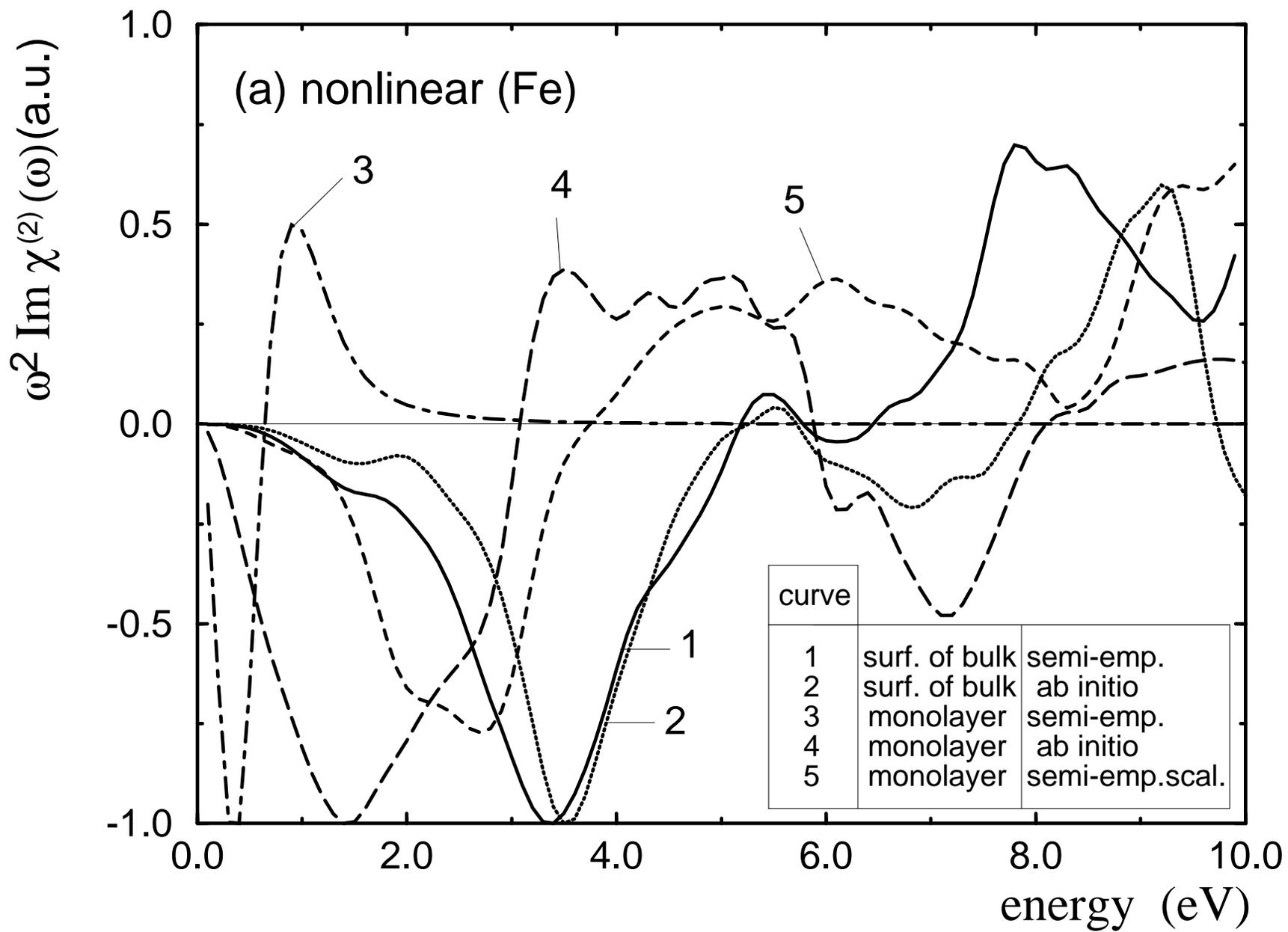
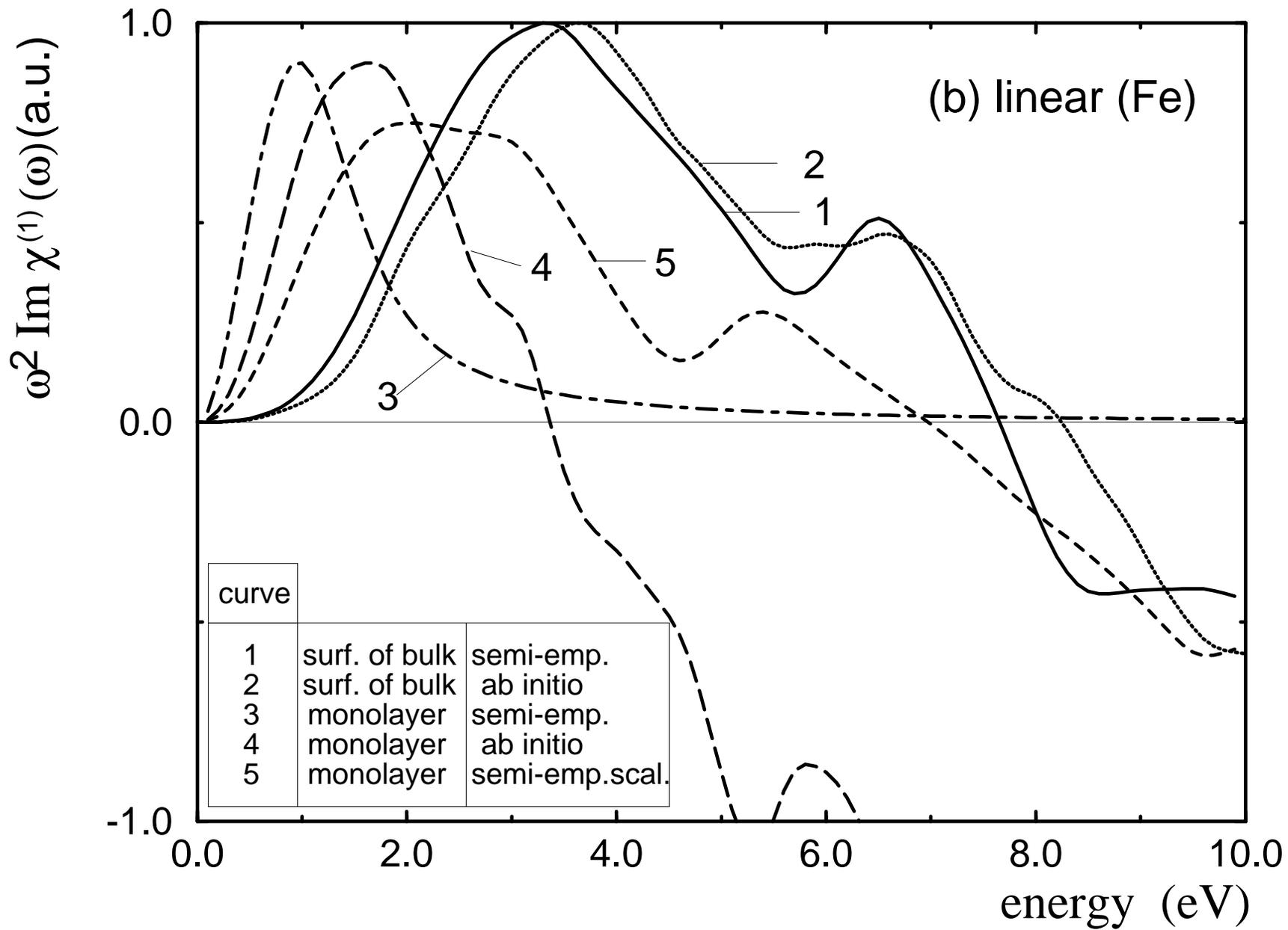
\begin{figure}
\caption[]{Comparison of {\em ab initio} and semi-empirical calculations
of the (a) nonlinear and (b) linear magneto-optical susceptibilities of Fe.
The bulk surface results from truncating bulk and the semi-empirical
calculations for a monolayer are performed with reduced hopping
parameters $A_{0}$.}
\end{figure}
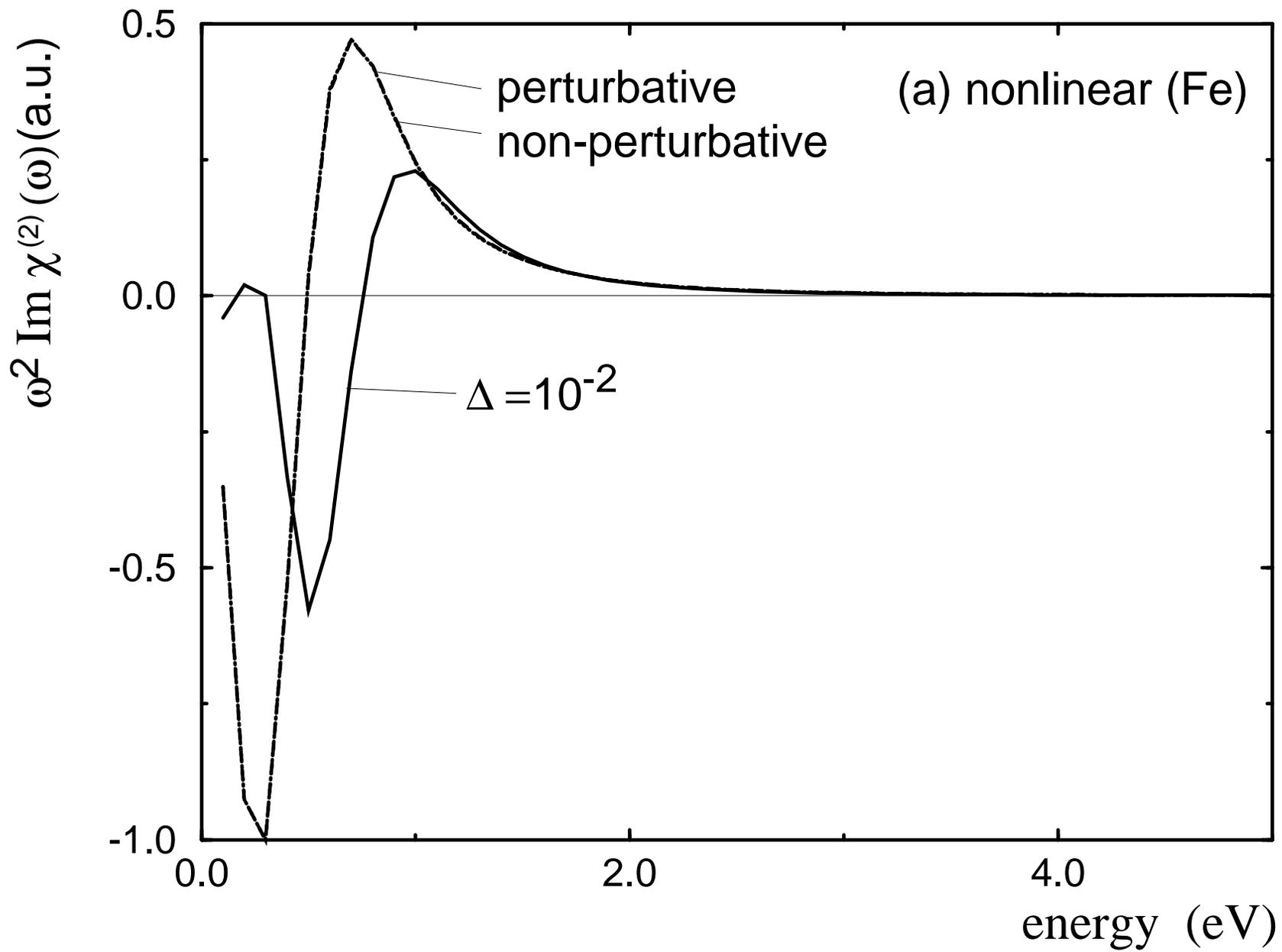
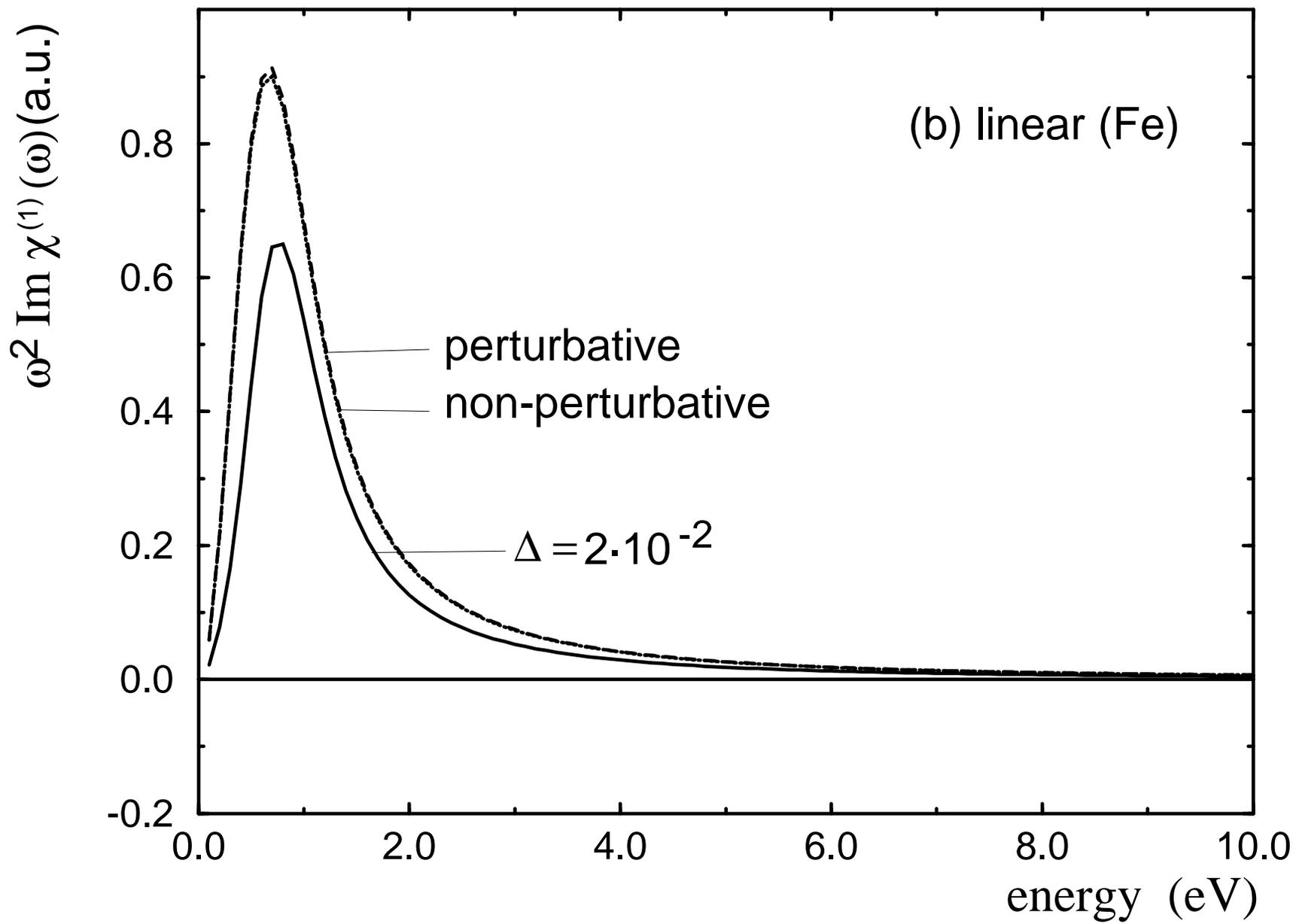
\begin{figure}
\caption[]{Role of spin-orbit coupling: Semiempirically calculated
(a) nonlinear and (b) linear magneto-optical Kerr spectra of a Fe monolayer
with spin-orbit coupling in first order perturbation theory (dotted curve)
and treated non-perturbatively (dashed curve).
$\Delta $ gives the difference.}
\end{figure}

\begin{references}
\bibitem[23]
{fuss} This can also be seen from first principle calculations treating
properly the matrix elements in $\chi^{(2)}$ which sense the breakdown of
inversion symmetry which is necessary for a SH signal. Such calculations
are in progress; U. Pustogowa {\em et al.}, to be published.
\end{references}
\end{document}